\documentclass[useAMS,usenatbib,usegraphicx,onecolumn]{mn2e}

\title[An updated IGM attenuation model]{An updated analytic model for
the attenuation by the intergalactic medium}
\author[A.\ K.\ Inoue et al.]{\parbox[t]{\textwidth}{\vspace{-1cm}
Akio K. Inoue$^{1}$\thanks{E-mail: akinoue@las.osaka-sandai.ac.jp},
Ikkoh Shimizu,$^{1,2}$, Ikuru Iwata$^{3}$, and Masayuki Tanaka$^{4}$}\\
$^{1}$College of General Education, Osaka Sangyo University, 
3-1-1, Nakagaito, Daito, Osaka 574-8530, Japan\\
$^{2}$Department of Astronomy, The University of Tokyo, 7-3-1
Hongo, Tokyo 113-0033, Japan\\
$^{3}$Subaru Telescope, National Astronomical Observatory of Japan, 
650 North A'ohoku Place Hilo, HI 96720, USA\\
$^{4}$National Astronomical Observatory of Japan, 
2-21-1, Osawa, Mitaka, Tokyo 181-8588, Japan}

\begin{document}

\date{}

\pagerange{\pageref{firstpage}--\pageref{lastpage}} \pubyear{2013}

\maketitle

\label{firstpage}

\begin{abstract}
 We present an updated version of the so-called Madau model for the
 attenuation by the intergalactic neutral hydrogen against the radiation
 from distant objects. First, we derive a distribution function of 
 the intergalactic absorbers from the latest observational statistics of
 the Ly$\alpha$ forest, Lyman limit systems, and damped Ly$\alpha$
 systems. The distribution function excellently reproduces the observed
 redshift evolutions of the Ly$\alpha$ depression and of the
 mean-free-path of the Lyman continuum simultaneously. Then, we derive a
 set of the analytic functions which describe the mean intergalactic
 attenuation curve for objects at $z>0.5$. The new model predicts 
 less (or more) Ly$\alpha$ attenuation for $z\simeq3$--5 ($z>6$)
 sources through usual broad-band filters relative to the original Madau
 model. This may cause a systematic difference in the photometric
 redshift estimates, which is, however, still as small as about 0.05.
 Finally, we find a more than 0.5 mag overestimation of the Lyman
 continuum attenuation in the original Madau model at $z>3$, which
 causes a significant overcorrection against direct observations of the
 Lyman continuum of galaxies.
\end{abstract}


\section{Introduction}

Radiation from cosmological sources is absorbed by neutral hydrogen left
in the intergalactic medium (IGM) even after the cosmic reionization
\citep[e.g.][]{gun65}. This intergalactic neutral hydrogen probably
traces `cosmic web' produced by the gravity of the dark matter
\citep[e.g.][]{rau98}. Along an observer's line-of-sight piercing
the cosmic web, there appear to be numerous discrete systems composed of
the intergalactic neutral hydrogen producing a number of absorption
lines in the spectra of distant sources. These systems are divided into
the Ly$\alpha$ forest (LAF; $\log_{10}(N_{\rm HI}/{\rm cm}^{-2})<17.2$), 
Lyman limit systems (LLSs; $17.2\leq\log_{10}(N_{\rm HI}/{\rm cm}^{-2})<20.3$) 
and damped Ly$\alpha$ systems (DLAs; $\log_{10}(N_{\rm HI}/{\rm cm}^{-2})\geq20.3$),
depending on the column density of the neutral hydrogen along the
line-of-sight \citep[e.g.][]{rau98}.

The intergalactic absorption is routinely found in the spectra of
objects at a cosmological distance and this feature is utilized as a
tool to select high-$z$ objects only by photometric data: the so-called
drop-out technique \citep[e.g.][]{ste95,mad96}. To put it another way,
we must always correct the spectra of cosmological sources for this
absorption in order to know the intrinsic ones. Therefore, an accurate
model of this absorption is quite useful as a standard tool for
observational cosmology.

After several models for this purpose were presented 
\citep[e.g.,][]{mol90,zuo93,yos94}, \cite{mad95} (hereafter
M95) appeared and became the most popular because of the convenient
analytic functions. However, the heart of the model, i.e. the statistics
of the LAF, LLSs, and DLAs, has been updated largely by observations in
the last two decades since M95. In fact, there are several papers
adopting such updated statistics \citep{ber99,mei06,tep08,ino08}. 
Nevertheless, people still adhere to M95, except for a few innovative
authors \citep[e.g.][]{har11,ove13}. This adherence may be due to the
simplicity and convenience of the analytic functions in M95. Here, in
this paper, we intend to present a user-friendly analytic function
conforming to updated statistics.
 
In the next section, we introduce the heart of the modeling: the
distribution function of the intergalactic absorbers derived from the
latest observational data of the LAF, LLSs, and DLAs. Then, we show the
updated mean transmission function and compare it with the latest
observations of the Ly$\alpha$ transmission and the mean-free-path of
Lyman limit photons in section 3. In section 4, we present new analytic
formulae of the intergalactic attenuation. Finally, we quantify the
difference of the attenuation magnitudes through some broad-band filters
between the M95 model and ours and discuss the effect on the drop-out
technique and the photometric redshift estimation in section 5. 
 We do not need to assume any specific cosmological model in
this paper, except for sections 2.1 and 3.2, where we assume 
$\Omega_{\rm M}=0.3$, $\Omega_{\Lambda}=0.7$, and $H_0=70$ km s$^{-1}$
Mpc$^{-1}$. Therefore, the analytic functions presented in section 4 can
be used directly in any cosmological models.

\section{Distribution function of intergalactic absorbers}

The mean optical depth at the observed wavelength $\lambda_{\rm obs}$
along a line-of-sight, where we assume that absorbers distribute
randomly\footnote{In fact, the absorbers are correlated with each other
and even with the observing distant object
\citep[e.g.,][]{slo11,rud13,pro14}. However, we reserve constructing a
model with such a correlation for future work as done in the
literature.}, towards a source at $z_{\rm S}$ is
\citep[e.g.][]{par80,mad95}  
\begin{equation}
 \langle \tau^{\rm IGM}_{\lambda_{\rm obs}}(z_{\rm S}) \rangle 
  = \int_0^{z_{\rm S}} \int_0^\infty 
  \frac{\partial^2 n}{\partial z \partial N_{\rm HI}}
  (1-e^{-\tau_{\rm abs}}) dN_{\rm HI} dz\,,
\end{equation}
where $\partial^2 n/\partial z/\partial N_{\rm HI}$ is the distribution
function of the intergalactic absorbers and 
$\tau_{\rm abs}=\sigma^{\rm HI}_{\lambda_{\rm abs}}N_{\rm HI}$ is the
optical depth of an absorber with the H {\sc i} column density 
$N_{\rm HI}$ at the redshift $z$ and the H {\sc i} cross section  
$\sigma^{\rm HI}_{\lambda_{\rm abs}}$, which includes an assumed
line profile for Lyman series absorption, at the wavelength in the absorber's
rest-frame $\lambda_{\rm abs}=\lambda_{\rm obs}/(1+z)$. By specifying 
the distribution function, we can integrate equation (1) analytically if
possible or numerically. Thus, an appropriate distribution function is
essential for the model of the intergalactic absorption.

M95 assumed the following function:
\begin{equation}
 \frac{\partial^2 n}{\partial z \partial N_{\rm HI}}
  = \cases{
  2.4\times10^7 N_{\rm HI}^{-1.5} (1+z)^{2.46} & 
  ($2\times10^{12}<N_{\rm HI}/{\rm cm^{-2}}<1.59\times10^{17}$) \cr
  1.9\times10^8 N_{\rm HI}^{-1.5} (1+z)^{0.68} &
  ($1.59\times10^{17}<N_{\rm HI}/{\rm cm^{-2}}<2\times10^{20}$) \cr
  }\,,
\end{equation}
where the column density distribution is assumed to be a single
power-law with the index of $-1.5$ but the redshift distribution is
divided into two parts: one for the LAF and the other for LLSs. Since
the two categories evolve separately along the redshift in this assumed
function, there is a discontinuity point in the column density
distribution as seen in Figure~1 (a). More recent work by \cite{mei06}
also adopted such a separate treatment of the LAF and LLSs.

Our previous work, \cite{ino08} (hereafter II08), assumed the following
function: 
\begin{equation}
 \frac{\partial^2 n}{\partial z \partial N_{\rm HI}} 
  = f(z)g(N_{\rm HI})\,,
\end{equation}
where $f(z)$ and $g(N_{\rm HI})$ are the distribution functions in $z$
and $N_{\rm HI}$ spaces, respectively. The spirit of this formulation is
a universal column density distribution in any $z$ and a common redshift
evolution for all absorbers independent of $N_{\rm HI}$, producing no
discontinuity in the column density distribution (see Fig.~1 [b]). 
II08 assumed a double power-law of $N_{\rm HI}$ for $g(N_{\rm HI})$ and
a triple power-law of $(1+z)$ for $f(z)$.

As an extension of the II08 formalism, we here introduce the following
function composed of two components named as the LAF and DLA components,
respectively:  
\begin{equation}
 \frac{\partial^2 n}{\partial z \partial N_{\rm HI}} 
  = f_{\rm LAF}(z)g_{\rm LAF}(N_{\rm HI})
  + f_{\rm DLA}(z)g_{\rm DLA}(N_{\rm HI})\,.
\end{equation}
This is motivated by recent observations of the absorbers' statistics,
especially the discovery of almost no evolution of the DLA encounter
probability per unit `absorption length' \citep{pro09a} and the new
measurements of the mean-free-path of the Lyman limit photons
\citep{pro09b,ome13,fum13,wor14}. We have found that it is very
difficult to reproduce all the observed statistics simultaneously with
the II08  formulation, i.e. a single component model. On the other hand,
the two component model newly introduced in this paper can reproduce all
the observations very well as shown in the following sections. Note that
the two components both have all categories of absorbers as described
below but one dominates the other in the column density range which the
name indicates. The two components both significantly contribute to
LLSs. We also note that the formulation in this paper is not the unique
solution but an example description reproducing all the observational
statistics. In this sense, we do not determine the parameters in the
functions by any statistical test but do just by eye in comparisons with
observations below. Nevertheless, we may suggest that the success of
this separate treatment of the LAF and DLAs means different origins of
the two categories. That is, the LAF traces diffuse filaments of the
cosmic web not associated with halos and galaxies yet
\citep[e.g.,][]{cen94}, but DLAs are associated with materials in halos
and galaxies \citep[e.g.,][]{hae98}.

In this paper, we assume a column density distribution function
matching with the observed shape but still integrable analytically. 
One example of such functions is as follows:
\begin{equation}
 g_i(N_{\rm HI}) = B_i {N_{\rm HI}}^{-\beta_i} e^{-N_{\rm HI}/N_{\rm c}}\,,
\end{equation}
where the subscript $i={\rm LAF}$ or DLA, $\beta_i$ is the power-law
index for each component, $N_{\rm c}$ is the cut-off column density
assumed to be common to the two components, and $B_i$ is the
normalization determined by 
$\int_{N_{\rm l}}^{N_{\rm u}} g_i(N_{\rm HI})dN_{\rm HI}=1$
with the boundaries of $N_{\rm l}$ and $N_{\rm u}$ which are also 
assumed to be common to the two components. Thus, each component
has in fact all types of absorbers from the LAF to DLAs, but the LAF (or
DLA) component negligibly contributes to the DLA (LAF) number density as
shown in Figure~1 (c) and (d). We also note that the function $g_i$ is
still continuous outside of these boundaries and we integrate $g_i$ from
0 to $\infty$ in equation (1). Some analytic integrations of this
function are found in appendix.

The redshift distribution functions, $f_i(z)$, are assumed
to be broken power-laws of $(1+z)$ as follows: for the LAF component, 
\begin{equation}
 f_{\rm LAF}(z) = {\cal A}_{\rm LAF} \cases{
  \left(\frac{1+z}{1+z_{\rm LAF,1}}\right)^{\gamma_{\rm LAF,1}} &
  ($z<z_{\rm LAF,1}$) \cr
  \left(\frac{1+z}{1+z_{\rm LAF,1}}\right)^{\gamma_{\rm LAF,2}} &
  ($z_{\rm LAF,1}\leq z<z_{\rm LAF,2}$) \cr 
  \left(\frac{1+z_{\rm LAF,2}}{1+z_{\rm LAF,1}}\right)^{\gamma_{\rm LAF,2}} 
  \left(\frac{1+z}{1+z_{\rm LAF,2}}\right)^{\gamma_{\rm LAF,3}} &
  ($z_{\rm LAF,2}\leq z$) \cr
  }\,,
\end{equation}
and for the DLA component, 
\begin{equation}
 f_{\rm DLA}(z) = {\cal A}_{\rm DLA} \cases{
  \left(\frac{1+z}{1+z_{\rm DLA,1}}\right)^{\gamma_{\rm DLA,1}} &
  ($z<z_{\rm DLA,1}$) \cr
  \left(\frac{1+z}{1+z_{\rm DLA,1}}\right)^{\gamma_{\rm DLA,2}} &
  ($z_{\rm DLA,1}\leq z$) \cr
  }\,.
\end{equation}
The normalization of each component ${\cal A}_i$, where $i={\rm LAF}$ or
DLA, is the number of absorbers with the column density 
$N_{\rm l} \leq N_{\rm HI} \leq N_{\rm u}$ per unit redshift interval at
the redshift $z_{i,1}$. Note that the normalization ${\cal A}_i$ would be
different if we chose other sets of $N_{\rm l}$ and $N_{\rm u}$. The
fiducial set of the parameters in the functions, $f_i$ and $g_i$, is
summarised in Table~1, which is obtained from comparisons with
observations as shown in the following sections.

\begin{table}
 \caption[]{Parameters for the distribution function of intergalactic
 absorbers assumed in this paper.}
 \setlength{\tabcolsep}{3pt}
 \footnotesize
 \begin{minipage}{\linewidth}
  \begin{tabular}{lccccccc}
   \hline
   Common \\
   Parameter & $\log_{10} (N_{\rm l}/{\rm cm}^{-2})$ & 
   $\log_{10} (N_{\rm u}/{\rm cm}^{-2})$ & 
   $\log_{10} (N_{\rm c}/{\rm cm}^{-2})$ & 
   $\langle b/{\rm km~s}^{-1} \rangle$ \\
   Value & 12 & 23 & 21 & 28 \\
   \hline
   LAF component \\
   Parameter & ${\cal A}_{\rm LAF}$ & $\beta_{\rm LAF}$ & 
   $z_{\rm LAF,1}$ & $z_{\rm LAF,2}$ & $\gamma_{\rm LAF,1}$ & 
   $\gamma_{\rm LAF,2}$ & $\gamma_{\rm LAF,3}$ \\
   Value & 500 & 1.7 & 1.2 & 4.7 & 0.2 & 2.7 & 4.5 \\
   \hline
   DLA component \\
   Parameter & ${\cal A}_{\rm DLA}$ & $\beta_{\rm DLA}$ & 
   $z_{\rm DLA,1}$ & --- & $\gamma_{\rm DLA,1}$ & 
   $\gamma_{\rm DLA,2}$ & --- \\
   Value & 1.1 & 0.9 & 2.0 & --- & 1.0 & 2.0 & --- \\
   \hline
  \end{tabular}
 \end{minipage}
\end{table}%

\begin{figure*}
 \begin{center}
  \includegraphics[width=10cm]{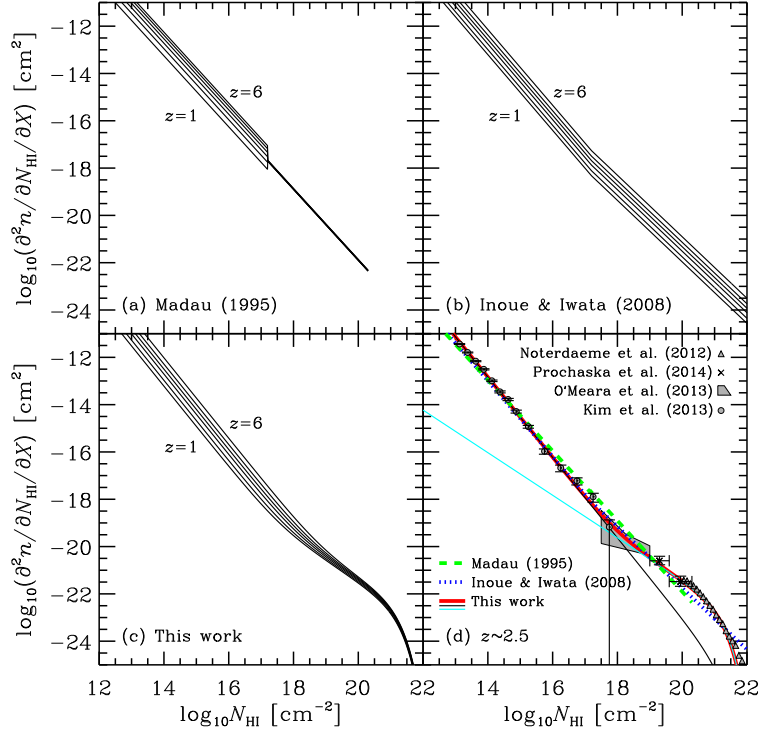}
 \end{center}
 \caption{Number of the intergalactic absorbers per unit column density
 of neutral hydrogen ($N_{\rm HI}$) per unit absorption length ($X$)
 along an average line of sight as a function of the column density. (a) 
 The redshift evolution of the functions in Madau (1995). (b) The same
 as (a) but for the model in Inoue \& Iwata (2008). (c) The same as (a)
 but for the model of this work. (d) A comparison of the three models
 with the observational data at $z\sim2.5$ taken from the literature:
 Kim et al.~(2013) for LAF, O'Meara et al.~(2013) for LLSs, Prochaska et
 al.~(2014) for sub-DLAs (original data presented by O'Meara et
 al.~2007), and Noterdaeme et al.~(2013) for DLAs. The solid, dotted,
 and dashed lines are models of this work, Inoue \& Iwata (2008), and
 Madau (1995), respectively. The two thin solid lines show the LAF and
 DLA components in this work.}
\end{figure*}

\subsection{Column density distribution}

First, we compare the column density distribution functions with
observations. The observed column density distributions are normally
described by the number of absorbers per unit column density 
$dN_{\rm HI}$ and per unit `absorption length' $dX$ \citep{bah69}: 
\begin{equation}
 \frac{\partial^2 n}{\partial X \partial N_{\rm HI}} 
  = \frac{\partial^2 n}{\partial z \partial N_{\rm HI}} \frac{dz}{dX}\,, 
\end{equation}
where 
\begin{equation}
 dX = \frac{H_0}{H(z)}(1+z)^2 dz\,,
\end{equation}
with the Hubble parameters $H_0$ at the current epoch and $H(z)$ at the
redshift $z$. Figure~1 (d) shows a comparison of the three models, this
work, II08, and M95, with the observations at $z\sim2.5$ compiled from
the literature. We see a good agreement between the observations and the
models. However, there are differences among them if we look closely as
discussed below.

M95 adopted a single power-law index of $-1.5$ \citep[e.g.][]{tyt87}.
However, recent observations suggest a break of the column density
distribution around $N_{\rm HI}\sim10^{17}$ cm$^{-2}$
\citep{pro05,pro10,ome13}; the slope changes from a steeper at a lower
column density to a shallower at a higher column density. The break
column density is about the threshold of LLSs; the optical depth against
the Lyman limit photon is about unity with this column density. 
Therefore, this break is probably caused by the transition between
optically thin and thick against the ionizing background radiation
\citep{cor01,cor02}. Indeed, the latest cosmological radiation
hydrodynamics simulations show that the self-shielding of the optically
thick absorbers is the mechanism producing the break
\citep{alt11,rah13}. On the other hand, M95 has the discontinuity at the
column density as found in Figure~1 (a) owing to the different redshift
evolutions of the LAF and LLSs.

II08 adopted a double power-law function in order to describe the break 
at $N_{\rm HI}\sim10^{17}$ cm$^{-2}$. In fact, they assumed the break
column density $N_{\rm HI}=1.6\times10^{17}$ cm$^{-2}$ at which the
optical depth against the Lyman limit photon becomes unity. With this
double power-law function for the column density distribution, II08
assumed a universal redshift evolution for all column densities. As a
result, the number densities of LLSs and DLAs per unit absorption length
monotonically increases with redshift as found in Figure~1 (b). However,
recent observations suggest a much weaker evolution of DLAs
\citep{pro05,ome13} and the cosmological simulations successfully
reproduce this weak evolution \citep{rah13}.

In this paper, we have assumed in equation (5) a power-law function with
an exponential cut-off like the Schechter function. Such a functional
shape has been already proposed by \cite{pro05} to describe the column
density distribution of DLAs. As found at highest column densities in
Figure~1 (d), this function reproduces the DLA distribution very well if
we adopt a cut-off column density $N_{\rm c}\simeq10^{21}$ cm$^{-2}$.
With this functional shape, we successfully reproduce the weak evolution
of the column density distribution of DLAs as shown in Figure~1 (c). This 
is partly due to a weaker redshift evolution in the DLA regime of this
paper than II08 as discussed in Figure~2 below, but the constancy of the
cut-off column density is also important. On the other hand, M95 also
predicts almost no evolution of LLSs due to its weak redshift evolution
for absorbers with the high column density (see eq.~[2]). However, the
M95 model does not have any DLAs.

\subsection{Number density evolution}

Next, we look into the number density evolution along the redshift. 
Figure~2 shows the comparison of the models with observations for four
categories of absorbers depending on the column density: 
$\log_{10}(N_{\rm HI}/{\rm cm}^{-2})>13.64$ (LAF), $>17.2$ (LLSs), 
$>19.0$ (sub-DLAs or super-LLSs), and $>20.3$ (DLAs). Figure~3 shows a
close-up of LLSs' evolution.

\begin{figure}
 \begin{center}
  \includegraphics[width=6cm]{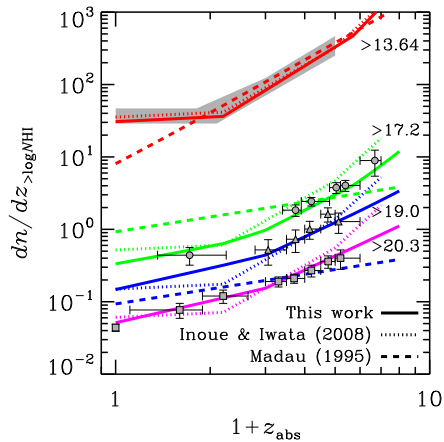}
 \end{center}
 \caption{Number of the intergalactic absorbers per unit redshift along
 an average line of sight as a function of the absorbers' redshift. The
 shaded area is the observed range for absorbers with 
 $\log_{10}(N_{\rm HI}/{\rm cm}^{-2})>13.6$ (LAF) taken from Weymann
 et al.~(1998), Kim et al.~(2001) and Janknecht et al.~(2006).  The
 filled circles, triangles and squares are the observed data of
 absorbers with  $\log_{10}(N_{\rm HI}/{\rm cm}^{-2})>17.2$ (LLS) taken
 from Songaila \& Cowie~(2010), $>19.0$ (sub-DLA) taken from P{\'e}roux
 et al.~(2005), and $>20.3$ (DLA) taken from Rao et al.~(2006),
 respectively. The solid, dotted, and dashed lines are the models of
 this work, Inoue \& Iwata (2008), and Madau (1995). Note that Madau
 (1995) model does not have DLAs.}
\end{figure}

\begin{figure*}
 \begin{center}
  \includegraphics[width=10cm]{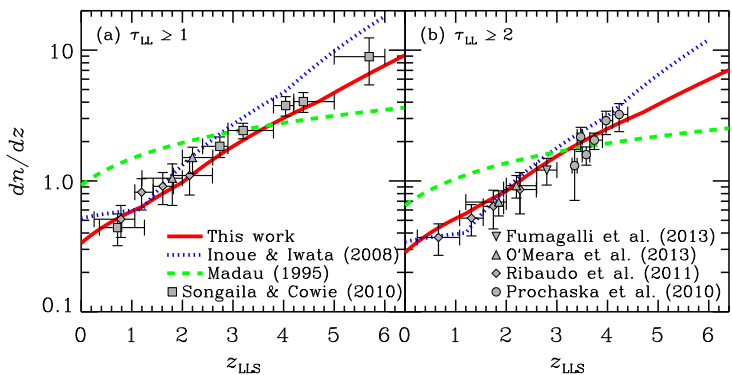}
 \end{center}
 \caption{Number of LLSs per unit redshift along an average line of
 sight as a function of the LLSs' redshift: (a) the systems with the
 optical depth for hydrogen Lyman limit photons equal to or larger
 than unity, and (b) the systems with the optical depth equal to or
 larger than two. The squares, diamonds, triangles, upside-down
 triangles, and circles are the observed data taken from Songaila \&
 Cowie (2010), Ribaudo et al.~(2011), O'Meara et al.~(2013),  Fumagalli
 et al.~(2013), and Prochaska et al.~(2010), respectively. 
 The solid, dotted, and dashed lines are the models of this work, Inoue
 \& Iwata (2008), and Madau (1995), respectively.}
\end{figure*}

M95 adopted a single power-law of $(1+z)$. This fits well with the LAF
number density evolution at $z>1$, but it predicts too small number
density relative to the observations at $z<1$, where the observed number
density is almost constant \citep{wey98}. The break of the observed
LAF number density evolution at $z\sim1$ is probably caused by the
sharp decline of the ionizing background radiation from the epoch to the
present \cite[e.g.][]{dav99}. The LLS number evolution of M95 is largely
different from the observations in respect of the slope, while the
absolute value matches the observations at $z\sim3$. Furthermore, the
M95 model has too small number of sub-DLAs and does not have any DLAs.

II08 adopted a twice-broken power-law for the redshift evolution. The
first break is set at $z\sim1$ to describe the bent of the LAF number
evolution and the second break is set at $z\sim4$ to reproduce a rapid
increase of the Ly$\alpha$ optical depth toward high-$z$ (see the next
section). The same function was assumed for all absorber categories in
II08. It is still compatible with the observed LLS and sub-DLA
evolutions, but the agreement becomes marginal for DLAs.

In this paper, we adopt two different evolutions for the LAF and DLA
components. As found in Figure~2, this new description shows the best
agreement with the observations for all absorber categories. Figure~3
shows that the new model matches with observations better than the II08
model. In particular, the new model tends to have a smaller number of
LLSs than II08. This point is essential to reproduce the observed
mean-free-path for ionizing photons as discussed in section 3.2.

\section{Mean transmission function}

With the distribution function of the intergalactic absorbers described
in the previous section, we can integrate equation (1) numerically and
obtain the mean transmission function of the IGM. In the integration, we
treat the neutral hydrogen cross section 
$\sigma^{\rm HI}_{\lambda_{\rm abs}}$ as follows; we adopt the
interpolation formula given by \cite{ost89} for the photoionization 
cross section. We also adopt the oscillator strengths and the damping
constants taken from \cite{wie66} and the analytic formula of the Voigt
profile given by \cite{tep06} for the Lyman series cross sections. The
mean Doppler velocity is assumed to be $\langle b \rangle=28$ km
s$^{-1}$ obtained from the $b$ distribution function proposed by
\cite{hui99} and its parameter $b_\sigma=23$ km s$^{-1}$ measured by
\cite{jan06} for this work (see Table~1) and II08. On the other hand, we
adopt $\langle b \rangle=35$ km s$^{-1}$ for M95 according to the
original assumption. In the integration of equation (1), we should set
the redshift step, $\Delta z$, to be fine enough to resolve the narrow
width of the Lyman series lines. We adopt $\Delta z=5\times10^{-5}$ and
have confirmed the convergence of the calculations.

\begin{figure}
 \begin{center}
  \includegraphics[width=7cm]{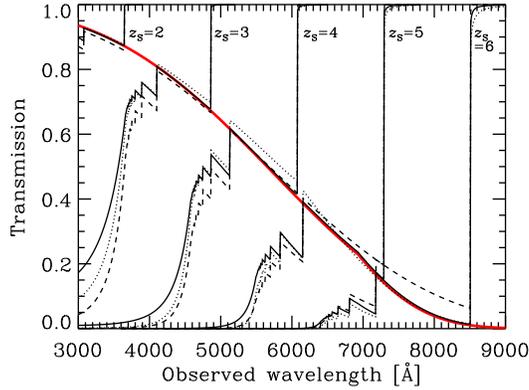}
 \end{center}
 \caption{Mean transmission functions for sources at $z_{\rm S}=2$ to
 6. The solid, dotted, and dashed lines are the models of this work,
 Inoue \& Iwata (2008), and Madau (1995), respectively. The thick solid
 line is an analytic approximation for Ly$\alpha$ transmission given by
 equation (19) for the model of this work.}
\end{figure}

Figure~4 shows the mean transmission functions obtained. The three
models are very similar but some differences are recognised if we look
at them in detail. In the regime of the Lyman series transmission for
$z_{\rm S}\leq4$, the II08 model is the highest, the M95 model is the
lowest, and the new model of this paper is middle. On the other hand,
the M95 model predicts the highest transmission for $z_{\rm S}\geq5$. 
However, the difference is small, except for the case of $z_{\rm S}=6$. 
This small difference comes from the small difference of the number
density of the LAF, which mainly produces the Lyman series absorption,
among the three models as seen in Figure~2. The deviation of the M95
model found in the wavelength between Ly$\alpha$ and Ly$\beta$ for 
$z_{\rm S}=6$ is due to the lack of a rapid increase of the LAF number
density at high-$z$ which are adopted in the other two models. This
point will be discussed again in Figure~5 below.

In the Lyman continuum regime, the new model predicts the highest
transmission, while the M95 model is the lowest for $z_{\rm S}\leq4$ and
the II08 model is the lowest for $z_{\rm S}\geq5$ (but it is difficult
to see in Fig.~4). Given that LLSs are mainly responsible for the Lyman
continuum absorption, this difference is caused by the difference of the
number density of LLSs. Indeed, M95 has the largest density of LLSs at 
$z_{\rm LLS}<3$ and the LLS density of II08 becomes the largest at 
$z_{\rm LLS}>3$ (see Fig.~3). In the next two subsections, we compare
the model transmissions with the observations more in detail in terms of
the Ly$\alpha$ transmission and the mean-free-path for ionizing photons.

\subsection{Ly$\alpha$ transmission}

\begin{figure}
 \begin{center}
  \includegraphics[width=6cm]{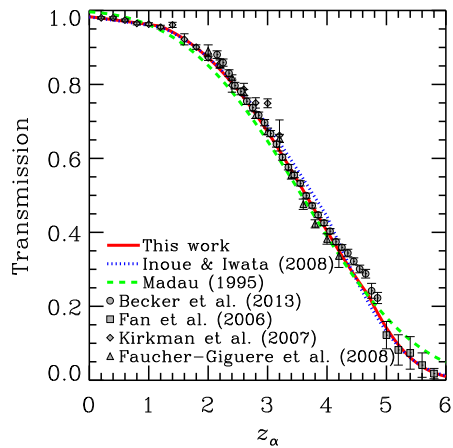}
 \end{center}
 \caption{Ly$\alpha$ transmission as a function of the redshift of
 Ly$\alpha$ line. The triangles, diamonds, squares, and circles are the
 observed data taken from Fauche-Giguere et al.~(2008), Kirkman et
 al.~(2007), Fan et al.~(2006), and Becker et al.~(2013),
 respectively. The solid, dotted, and dashed lines are the models of
 this work, Inoue \& Iwata (2008), and Madau (1995), respectively.}
\end{figure}

The spectrum between Ly$\alpha$ and Ly$\beta$ lines in the source
rest-frame is absorbed by the Ly$\alpha$ transition of the neutral
hydrogen in the IGM. This is called the Ly$\alpha$ depression (DA). Here
we compare the three models discussed in this paper with the
measurements of $1-$DA, i.e. Ly$\alpha$ transmission ($T_\alpha$), in
Figure~5. M95 presented an analytic formula for the mean optical depth
which corresponds to the transmission as 
$T_\alpha = \exp\left[-3.6\times10^{-3}(1+z_\alpha)^{3.46}\right]$, 
where $z_\alpha=\lambda_{\rm obs}/\lambda_\alpha-1$ is the redshift of
absorbers and the Ly$\alpha$ wavelength $\lambda_\alpha=1215.67$ \AA.
This formula is shown by the dashed line in Figure~5. We have confirmed
this formula from the mean transmission function which we obtained from
the integration of equation (1) with the M95 distribution function
(eq.~[2]).\footnote{There is a small difference ($<10\%$) between the
M95 $T_\alpha$ formula and that obtained from our numerical integration
of equation~(1). This is probably because the M95 formula was obtained
from another distribution function based on the equivalent width, not
based on the column density which we adopted in this paper.} We derive
analytic formulae of $T_\alpha$ from the mean transmission functions
of the II08 model and our new model which are shown by the dotted and
the solid lines in Figure~5, respectively. The derivation is given in
section 4 below. We find that all the three models excellently agree
with the observations. Among them, the new model presented in this paper
seems the best. In fact, the parameters for the LAF number density
evolution along the redshift of the new model (eq.~[6]) were chosen so
as to match the observed $T_\alpha$. However, the agreement is not
perfect at $z\simeq4.6$ where the observed data deviates upwards from
the model. Although we found a better agreement with the data if we
adopted a triple-break model instead of the double-break as equation
(6), we avoid it to keep the model as simple as possible. If further
observations emphasise the deviation, we should update the model again
in the future. On the other hand, the M95 model predicts slightly smaller
$T_\alpha$ at $1<z<4$ and larger at other redshifts than the observed
data. In particular, the M95 model deviates from the observations at
$z>5$ because it does not have a rapid increase of the LAF number
density towards the epoch of the reionization found in the last decade
\citep[e.g.][]{fan06}.

\subsection{Mean-free-path of ionizing photons}

\cite{pro09b} (hereafter PWO) proposed a new method for measuring
the mean-free-path in the IGM for ionizing photons directly in a
composite spectrum of QSOs. Here we follow the method updated by
\cite{ome13} \citep[see also][]{fum13,wor14}. Suppose a source at the
redshift $z_{\rm S}$ emitting an ionizing photon of the wavelength
$\lambda_{\rm S}<\lambda_{\rm L}$, where the Lyman limit wavelength
$\lambda_{\rm L}=911.8$ \AA\ \citep{cox00}. The wavelength of the photon
is redshifted cosmologically as it travels through the IGM. We suppose
then that it becomes $\lambda_{\rm L}$ at the redshift $z_{\rm L}$. 
Namely, $\lambda_{\rm L}(1+z_{\rm L})=\lambda_{\rm S}(1+z_{\rm S})$. The
optical depth between $z_{\rm S}$ and $z_{\rm L}$ along the light path
can be expressed as \citep{ome13}\footnote{PWO and \cite{ome13} started
from the following definition similar to equation (1) but a different
interval in the redshift integration: 
$$ \tau(z_{\rm L},z_{\rm S}) = \int_{z_{\rm L}}^{z_{\rm S}} \int_0^\infty 
  \frac{\partial^2 n}{\partial z \partial N_{\rm HI}}
  \left(1-e^{-N_{\rm HI}\sigma_{\rm ph}(z)}\right) dN_{\rm HI} dz\,,$$
where $\sigma_{\rm ph}(z)=\sigma_{\rm L}(1+z/1+z_{\rm L})^\alpha$ with
$\sigma_{\rm L}$ and $\alpha$ being the Lyman limit cross section of
neutral hydrogen and the power-law index of the wavelength dependence of
the cross section, respectively. Then, they introduced the opacity
$\kappa_{\rm L}(z)$ approximately to express the column density
integration. Although the exact expression of the column density
integration for the absorbers' function assumed in this paper is found
in the appendix, we keep their expression to compare their measurements with
our calculations directly.}
\begin{equation}
 \tau(z_{\rm L},z_{\rm S}) = \int_{z_{\rm L}}^{z_{\rm S}}
  \kappa_{\rm L}(z)\left(\frac{1+z}{1+z_{\rm L}}\right)^{-2.75}
  \left|\frac{dl}{dz}\right| dz\,,
\end{equation}
where $\kappa_{\rm L}$ is the IGM opacity for ionizing photons and 
$|dl/dz|$ is the proper length element per redshift. The index $-2.75$
comes from an approximate cross section of hydrogen atom for ionizing
photons \citep{ome13}. Assuming  
$\kappa_{\rm L}(z)=\kappa_{\rm L,S}(1+z/1+z_{\rm S})^\gamma$
and $|dl/dz|\approx(c/H_0)\Omega_{\rm M}^{-1/2}(1+z)^{-3/2}$, we obtain 
\begin{equation}
 \tau(\lambda_{\rm S},z_{\rm S}) \approx \kappa_{\rm L,S} 
  \frac{c(\lambda_{\rm S}/\lambda_{\rm L})^{2.75}
  \{(\lambda_{\rm S}/\lambda_{\rm L})^{\gamma-4.25}-1\}}
  {H_0\Omega_{\rm M}^{1/2}(1+z_{\rm S})^{3/2}(4.25-\gamma)}\,,
\end{equation}
where we have replaced $1+z_{\rm L}$ by 
$(1+z_{\rm S})(\lambda_{\rm S}/\lambda_{\rm L})$. Then, the
mean-free-path at the redshift $z_{\rm S}$ is defined by
\citep{pro09b,ome13,fum13} 
\begin{equation}
 l_{\rm mfp}(z_{\rm S}) \equiv \frac{1}{\kappa_{\rm L,S}}\,.
\end{equation}
\cite{pro09b}, \cite{ome13}, \cite{fum13}, \cite{wor14} obtained
the pivot opacities $\kappa_{\rm L,S}$ by fitting their composite
spectra of QSOs at various $z_{\rm S}$ with a function of 
$f_{\lambda_{\rm S}}/f_{\rm L}=\exp(-\tau[\lambda_{\rm S},z_{\rm S}])$, 
where $f_{\lambda_{\rm S}}$ and $f_{\rm L}$ are the flux densities of
the composite spectra at the wavelength $\lambda_{\rm S}$ and {\bf at}
the Lyman limit in the source rest-frame, respectively. Here we make a
very similar fitting with the transmission functions numerically
obtained in the previous subsection, $T_{\lambda_{\rm S}}$, because we
can express the composite spectrum as 
$f_{\lambda_{\rm S}}=f_{\lambda_{\rm S}}^{\rm int}T_{\lambda_{\rm S}}$, 
assuming an intrinsic QSO spectrum, $f_{\lambda_{\rm S}}^{\rm int}$.
We here adopt the power-law spectrum reported by \cite{tel02} for 
$f_{\lambda_{\rm S}}^{\rm int}$, while the change of the power-law index
does not have a large impact because of the narrow wavelength range used
for the fitting. The index $\gamma$ in equation (11) is set to be 2.0 
for the new model of this paper ($=\gamma_{\rm DLA,2}$), 2.5 for the
II08 model, and 0.68 for the M95 model, according to the LLS number
density evolution at the most relevant redshift range. However, the
choice does not affect the results very much \citep{ome13,fum13}.

There is another definition of the mean-free-path, which may be more
straightforward than that described above but more theoretical. Starting
from equation (1), we can define the mean IGM opacity at the Lyman limit
at $z_{\rm abs}$ as
\begin{equation}
 \kappa_{\rm L}(z_{\rm abs}) \equiv  
  \frac{d\langle \tau^{\rm IGM}_{\lambda_{\rm L}}(z_{\rm abs}) \rangle}
  {dz} \left|\frac{dz}{dl}\right| 
  = \left|\frac{dz}{dl}\right| \int_0^\infty
  \frac{\partial^2 n}{\partial z \partial N_{\rm HI}}
  (1-e^{-\sigma_{\rm L}N_{\rm HI}}) dN_{\rm HI} \,,
\end{equation}
where we have replaced $\lambda_{\rm obs}$ by $\lambda_{\rm L}(1+z_{\rm S})$
and $z_{\rm S}$ by $z_{\rm abs}$. The latter replacement means that the
opacity of equation (13) is one for the Lyman limit photon in the
immediate proximity to the redshift emitted; $\lambda_{\rm abs}=\lambda_{\rm L}$. 
Note that $\sigma_{\rm L}$ is the photoionization cross section for the
Lyman limit photon. We can integrate equation (13) numerically easily
with the absorber distribution function assumed. Then, the
mean-free-path is given by
\begin{equation}
 l_{\rm mfp}(z_{\rm abs})\equiv \frac{1}{\kappa_{\rm L}(z_{\rm abs})}\,.
\end{equation}

\begin{figure*}
 \begin{center}
  \includegraphics[width=10cm]{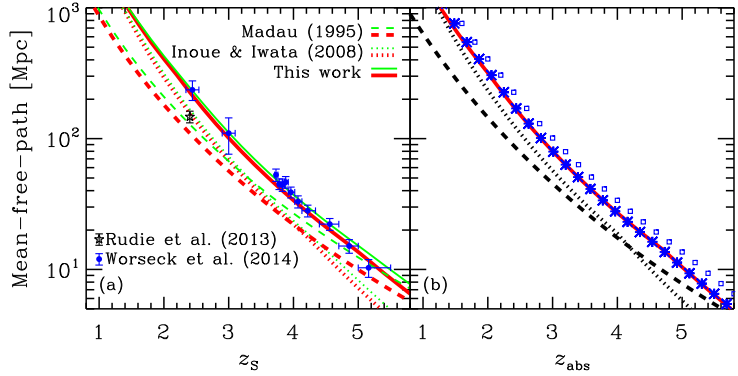}
 \end{center}
 \caption{Mean-free-path for hydrogen Lyman limit photons as a function
 of (a) the source redshift and (b) the absorbers' redshift. (a) The
 circles with error-bars are the data taken from a compilation of
 Worseck et al.~(2014). The solid, dotted, and dashed lines are the
 estimates obtained from the mean transmission functions of this paper,
 Inoue \& Iwata (2008), and Madau (1995), respectively, by using the
 PWO method (eqs.~[11] and [12]). The thick lines are the
 estimates within the wavelength range of 837 to 905 \AA\ in the source
 rest-frame, which should be compared with the data at $z\geq3$. The
 thin lines are that of 700 to 911 \AA, which should be compared with
 the data at $z\sim2.4$. (b) The solid, dotted, and dashed lines are the
 same as the panel (a), but based on equations (13) and (14). The small
 squares are the same result as the solid thick line in the panel (a)
 and these are moved to the asterisks by the conversion from the source
 redshift to the absorbers' redshift.}
\end{figure*}

Figure~6 shows the resultant mean-free-paths. In the panel (a), we show 
the comparison of the mean-free-path taken from \cite{wor14} with
the three models discussed in this paper. Their measurements are
obtained from the fitting in the wavelength range of 837--905 \AA\ for
$z\sim4$ \citep{pro09b}, 700--911 \AA\ for $z=2.4$ \citep{ome13}, 
830--905 \AA\ for $z=3$ \citep{fum13}, and 850--910 \AA\ for $z>4.5$
\citep{wor14}. These wavelength ranges are in the source rest-frame. We
have made fittings of the mean transmission curves in the $z=2.4$ and
$z\sim4$ wavelength ranges and obtained the mean-free-paths which are
shown by the solid, dotted, and dashed lines in the panel (a). We find
an excellent agreement between the new model and the observations. On
the other hand, the II08 and M95 models predict shorter mean-free-paths
than the observations. At this stage, these old models have been
inconsistent with the observations.

There is another recent measurement of the mean-free-path at
$z\approx2.4$ by \cite{rud13} which is a factor of 2 shorter than that by
\cite{ome13}. According to \cite{pro14}, the effects of line-blending
and clustering of strong absorption systems like LLSs may cause such a
discrepancy. In this paper, we adopt the measurement by \cite{ome13} for
$z\approx2.4$ and just show the measurement by \cite{rud13} in Figure~6
(a) for a comparison.

Figure~6 (b) shows the difference between the two definitions of the
mean-free-path introduced above (eqs.~[12] and [14]). The solid line is
the mean-free-path calculated from equation (14), but the small squares
are calculated from equation (12): the PWO method. We find
a small displacement between the two. We consider its origin to be the
difference of the redshifts; the displacement is horizontal not
vertical. The PWO method (eq.~[12]) measures the
mean-free-path of the photons with the wavelength $\approx870$ \AA\ at
$z_{\rm S}$. But the wavelength of this photon is redshifted to the
Lyman limit at $z_{\rm abs}$ at which equation (14) gives the
mean-free-path. If we convert $z_{\rm S}$ in the PWO
method into $z_{\rm abs}$, we obtain the asterisks and find an excellent
agreements of the two mean-free-paths. Therefore, one should take care
of the definition of the mean-free-path to compare a result with
another.

\section{New analytic model}

In this section, we derive a set of analytic approximation formulae
for the mean transmission function numerically obtained with the new
distribution function of absorbers. For an analytic integration of
equation (1), we approximate Lyman series line cross section profiles to
be a narrow rectangular shape. Then, we treat each line optical depth
and the Lyman continuum optical depth occurring at an observed
wavelength $\lambda_{\rm obs}$ separately. Namely,
\begin{equation}
 \langle \tau^{\rm IGM}_{\lambda_{\rm obs}}(z_{\rm S}) \rangle 
  \approx \tau_{\rm LS}^{\rm LAF}(\lambda_{\rm obs},z_{\rm S})
  + \tau_{\rm LS}^{\rm DLA}(\lambda_{\rm obs},z_{\rm S})
  + \tau_{\rm LC}^{\rm LAF}(\lambda_{\rm obs},z_{\rm S})
  + \tau_{\rm LC}^{\rm DLA}(\lambda_{\rm obs},z_{\rm S})\,.
\end{equation}
The Lyman series (LS) optical depths are given as 
\begin{equation}
 \tau_{\rm LS}^i(\lambda_{\rm obs},z_{\rm S}) 
  = \sum_j \tau^i_j(\lambda_{\rm obs},z_{\rm S})
  = \sum_j \int_0^{z_{\rm S}} f_i(z) I_{i,j}(z) dz  \,,
\end{equation}
where $\tau^i_j$ is the optical depth of $j$th line of Lyman series of 
$i$ (LAF or DLA) component and $I_{i,j}$ is the integration of the
column density function $g_i$ for the $j$th line given by equations
(A13) and (A14). Likewise, the Lyman continuum (LC) optical depths are
given as
\begin{equation}
 \tau_{\rm LC}^i(\lambda_{\rm obs},z_{\rm S})
  = \int_0^{z_{\rm S}} f_i(z) I_{i,{\rm LC}}(z) dz \,,
\end{equation}
where $I_{i,{\rm LC}}(z)$ is the column density integral again given by
equations (A13) and (A14). In the following subsections, we present 
analytic formulae of these optical depths.

\subsection{Lyman series absorption}

For the Lyman series absorption, let us assume a narrow rectangular
shape of the cross section of the Lyman series lines approximately in
the integration of equation (1). For example, the Ly$\alpha$ cross
section is assumed to be $\sigma_\alpha(\lambda)=\sigma_{\alpha,0}$ for 
$\lambda_\alpha-\Delta\lambda/2<\lambda<\lambda_\alpha+\Delta\lambda/2$
and 0 for otherwise, where $\sigma_{\alpha,0}$ is the cross section at
the line center of the Ly$\alpha$ wavelength $\lambda_\alpha$. The width
$\Delta\lambda$ can be expressed as $(\delta b/c) \lambda_\alpha$ where
$\delta$ is a numerical factor, $b$ is the Doppler velocity, and $c$ is
the light speed in the vacuum. If we assume a Gaussian line profile and
the cross section integrated over the wavelength from 0 to $\infty$ to
be equal to $\sigma_{\rm \alpha,0} \Delta\lambda$, we obtain 
$\delta=\sqrt{\pi}$. However, the DLA component may contribute to the
optical depth especially of higher order Lyman series lines, and in this
case, there may be a contribution from the damping wing to the
cross section, so that a larger $\delta$ value may be favorable. We thus
determine the values of $\delta$ from a comparison with the numerical
integration later.

In the rectangular cross section approximation, the optical depth for
the Ly$\alpha$ absorption at the observed wavelength  
$\lambda_{\rm obs}=\lambda_\alpha(1+z_\alpha)$ is produced by absorbers
within a narrow redshift range $(1+z_\alpha)\pm\Delta(1+z_\alpha)/2$, 
where $\Delta(1+z_\alpha)\approx(1+z_\alpha)(\delta b/c)$ if we omit the
term including $(\delta b/c)^2$. Then, the Ly$\alpha$ optical depth is
independent of the source redshift $z_{\rm S}$ but just depends on the
absorbers' redshift $z_\alpha$. We express it as
$\tau_\alpha(z_\alpha)$. For this Ly$\alpha$ optical depth, the
contribution of the DLA component can be neglected. In this case,
equation (1) becomes 
\begin{equation}
 \tau_\alpha(z_\alpha) \approx f_{\rm LAF}(z_\alpha) (1+z_\alpha) 
  \left(\frac{\delta b}{c}\right) \int_0^\infty g_{\rm LAF}(N_{\rm HI})
  (1-e^{-\sigma_{\alpha,0}N_{\rm HI}}) dN_{\rm HI}\,.
\end{equation}
According to the analytic integration of the column density distribution
presented in equation (A13), we obtain the Ly$\alpha$ optical depth as 
\begin{equation}
 \tau_\alpha(z_\alpha) \approx  f_{\rm LAF}(z_\alpha) (1+z_\alpha) 
  \left(\frac{\delta b}{c}\right) 
  (\sigma_{\alpha,0}N_{\rm l})^{\beta_{\rm LAF}-1} \Gamma(2-\beta_{\rm LAF})\,,
\end{equation}
where $\Gamma$ is a Gamma function. Adopting $\delta=\sqrt{\pi}$
(i.e., a Gaussian line profile approximation), we find an excellent
agreement with the numerical solution as shown in Figure~4. This indeed 
indicates that the Ly$\alpha$ transmission is determined almost only by
the LAF component. For other Lyman series lines, we replace 
$\sigma_{\alpha,0}$ with $\sigma_{j,0}$ for the $j$th line, and then,
the optical depth function for the $j$th line has the same functional
shape as that of Ly$\alpha$:
\begin{equation}
 \tau^{\rm LAF}_j(z_j) \propto f_{\rm LAF}(z_j)(1+z_j)\,,
\end{equation}
where $1+z_j=\lambda_{\rm obs}/\lambda_j$ with the wavelength of the
$j$th line $\lambda_j$. Given the functional shape of $f_{\rm LAF}$ in
equation (6) with the fiducial set of the parameters, we obtain, 
for $\lambda_j < \lambda_{\rm obs} < \lambda_j(1+z_{\rm S})$, 
\begin{equation}
 \tau_j^{\rm LAF}(\lambda_{\rm obs}) = \cases{
  A_{j,1}^{\rm LAF}\left(\frac{\lambda_{\rm obs}}{\lambda_j}\right)^{1.2} &
  ($\lambda_{\rm obs}<2.2\lambda_j$) \cr
  A_{j,2}^{\rm LAF}\left(\frac{\lambda_{\rm obs}}{\lambda_j}\right)^{3.7} &
  ($2.2\lambda_j\leq\lambda_{\rm obs}<5.7\lambda_j$) \cr 
  A_{j,3}^{\rm LAF}\left(\frac{\lambda_{\rm obs}}{\lambda_j}\right)^{5.5} &
  ($5.7\lambda_j\leq\lambda_{\rm obs}$) \cr
  }\,,
\end{equation}
otherwise $\tau_j^{\rm LAF}(\lambda_{\rm obs}) =0$.
The coefficients $A_{j,k}^{\rm LAF}$ ($k=1$, 2, and 3) calculated with
$\delta=\sqrt{\pi}$ are summarised in Table 2 with the wavelength
$\lambda_j$ up to the 40th line considered in this paper.

Comparing this analytic model with the numerical integration, we find a
slight difference between them at higher order lines. This is
qualitatively because the contribution of the DLA component increases 
for higher order lines. Note that the DLA component in this paper still
has absorbers in the LLS and even LAF column density regimes. Therefore,
we also consider the contribution of the DLA component to the Lyman
series line absorption. This contribution is also expressed by equation
(20) but with the DLA number density function, $f_{\rm DLA}$. From 
the comparison with the numerical solution, we find a good fit if we set
$\delta=5.0$ for the DLA component. Likewise the LAF case, we express
the analytic optical depth as, for 
$\lambda_j < \lambda_{\rm obs} < \lambda_j(1+z_{\rm S})$, 
\begin{equation}
 \tau_j^{\rm DLA}(\lambda_{\rm obs}) = \cases{
  A_{j,1}^{\rm DLA}\left(\frac{\lambda_{\rm obs}}{\lambda_j}\right)^{2.0} &
  ($\lambda_{\rm obs}<3.0\lambda_j$) \cr
  A_{j,2}^{\rm DLA}\left(\frac{\lambda_{\rm obs}}{\lambda_j}\right)^{3.0} &
  ($\lambda_{\rm obs}\geq3.0\lambda_j$) \cr 
  }\,,
\end{equation}
otherwise $\tau_j^{\rm DLA}(\lambda_{\rm obs}) =0$.
Table 2 presents a list of the coefficients.

\begin{table*}
 \caption[]{Wavelengths and coefficients for Lyman series absorption.}
 \setlength{\tabcolsep}{3pt}
 \footnotesize
 \begin{minipage}{\linewidth}
  \begin{tabular}{lcccccc}
   \hline
   $j$ & $\lambda_j$ (\AA) & $A_{j,1}^{\rm LAF}$ & $A_{j,2}^{\rm LAF}$ 
   & $A_{j,3}^{\rm LAF}$ & $A_{j,1}^{\rm DLA}$ & $A_{j,2}^{\rm DLA}$ \\
   \hline
   2 (Ly$\alpha$) & 1215.67 & 1.690e-02 & 2.354e-03 & 1.026e-04 & 1.617e-04 & 5.390e-05 \\
   3 (Ly$\beta$) & 1025.72 & 4.692e-03 & 6.536e-04 & 2.849e-05 & 1.545e-04 & 5.151e-05 \\
   4 (Ly$\gamma$) &  972.537 & 2.239e-03 & 3.119e-04 & 1.360e-05 & 1.498e-04 & 4.992e-05 \\
   5 &  949.743 & 1.319e-03 & 1.837e-04 & 8.010e-06 & 1.460e-04 & 4.868e-05 \\
   6 &  937.803 & 8.707e-04 & 1.213e-04 & 5.287e-06 & 1.429e-04 & 4.763e-05 \\
   7 &  930.748 & 6.178e-04 & 8.606e-05 & 3.752e-06 & 1.402e-04 & 4.672e-05 \\
   8 &  926.226 & 4.609e-04 & 6.421e-05 & 2.799e-06 & 1.377e-04 & 4.590e-05 \\
   9 &  923.150 & 3.569e-04 & 4.971e-05 & 2.167e-06 & 1.355e-04 & 4.516e-05 \\
   10 &  920.963 & 2.843e-04 & 3.960e-05 & 1.726e-06 & 1.335e-04 & 4.448e-05 \\
   11 &  919.352 & 2.318e-04 & 3.229e-05 & 1.407e-06 & 1.316e-04 & 4.385e-05 \\
   12 &  918.129 & 1.923e-04 & 2.679e-05 & 1.168e-06 & 1.298e-04 & 4.326e-05 \\
   13 &  917.181 & 1.622e-04 & 2.259e-05 & 9.847e-07 & 1.281e-04 & 4.271e-05 \\
   14 &  916.429 & 1.385e-04 & 1.929e-05 & 8.410e-07 & 1.265e-04 & 4.218e-05 \\
   15 &  915.824 & 1.196e-04 & 1.666e-05 & 7.263e-07 & 1.250e-04 & 4.168e-05 \\
   16 &  915.329 & 1.043e-04 & 1.453e-05 & 6.334e-07 & 1.236e-04 & 4.120e-05 \\
   17 &  914.919 & 9.174e-05 & 1.278e-05 & 5.571e-07 & 1.222e-04 & 4.075e-05 \\
   18 &  914.576 & 8.128e-05 & 1.132e-05 & 4.936e-07 & 1.209e-04 & 4.031e-05 \\
   19 &  914.286 & 7.251e-05 & 1.010e-05 & 4.403e-07 & 1.197e-04 & 3.989e-05 \\
   20 &  914.039 & 6.505e-05 & 9.062e-06 & 3.950e-07 & 1.185e-04 & 3.949e-05 \\
   21 &  913.826 & 5.868e-05 & 8.174e-06 & 3.563e-07 & 1.173e-04 & 3.910e-05 \\
   22 &  913.641 & 5.319e-05 & 7.409e-06 & 3.230e-07 & 1.162e-04 & 3.872e-05 \\
   23 &  913.480 & 4.843e-05 & 6.746e-06 & 2.941e-07 & 1.151e-04 & 3.836e-05 \\
   24 &  913.339 & 4.427e-05 & 6.167e-06 & 2.689e-07 & 1.140e-04 & 3.800e-05 \\
   25 &  913.215 & 4.063e-05 & 5.660e-06 & 2.467e-07 & 1.130e-04 & 3.766e-05 \\
   26 &  913.104 & 3.738e-05 & 5.207e-06 & 2.270e-07 & 1.120e-04 & 3.732e-05 \\
   27 &  913.006 & 3.454e-05 & 4.811e-06 & 2.097e-07 & 1.110e-04 & 3.700e-05 \\
   28 &  912.918 & 3.199e-05 & 4.456e-06 & 1.943e-07 & 1.101e-04 & 3.668e-05 \\
   29 &  912.839 & 2.971e-05 & 4.139e-06 & 1.804e-07 & 1.091e-04 & 3.637e-05 \\
   30 &  912.768 & 2.766e-05 & 3.853e-06 & 1.680e-07 & 1.082e-04 & 3.607e-05 \\
   31 &  912.703 & 2.582e-05 & 3.596e-06 & 1.568e-07 & 1.073e-04 & 3.578e-05 \\
   32 &  912.645 & 2.415e-05 & 3.364e-06 & 1.466e-07 & 1.065e-04 & 3.549e-05 \\
   33 &  912.592 & 2.263e-05 & 3.153e-06 & 1.375e-07 & 1.056e-04 & 3.521e-05 \\
   34 &  912.543 & 2.126e-05 & 2.961e-06 & 1.291e-07 & 1.048e-04 & 3.493e-05 \\
   35 &  912.499 & 2.000e-05 & 2.785e-06 & 1.214e-07 & 1.040e-04 & 3.466e-05 \\
   36 &  912.458 & 1.885e-05 & 2.625e-06 & 1.145e-07 & 1.032e-04 & 3.440e-05 \\
   37 &  912.420 & 1.779e-05 & 2.479e-06 & 1.080e-07 & 1.024e-04 & 3.414e-05 \\
   38 &  912.385 & 1.682e-05 & 2.343e-06 & 1.022e-07 & 1.017e-04 & 3.389e-05 \\
   39 &  912.353 & 1.593e-05 & 2.219e-06 & 9.673e-08 & 1.009e-04 & 3.364e-05 \\
   40 &  912.324 & 1.510e-05 & 2.103e-06 & 9.169e-08 & 1.002e-04 & 3.339e-05 \\
   \hline
  \end{tabular}
 \end{minipage}
\end{table*}%

\subsection{Lyman continuum absorption}

Substituting equations (A13) and (A14) for the column density integral,  
$I_{i,{\rm LC}}(z)$, in equation (17), for the LAF and DLA components,
respectively, we obtain the optical depths by the two components as 
\begin{equation}
 \tau^{\rm LAF}_{\rm LC}(\lambda_{\rm obs},z_{\rm S})
  \approx \Gamma(2-\beta_{\rm LAF}) 
  (N_{\rm l}\sigma_{\rm L})^{\beta_{\rm LAF}-1}
  \int_0^{z_{\rm S}} f_{\rm LAF}(z)
  \left(\frac{1+z_{\rm L}}{1+z}\right)^{\alpha(\beta_{\rm LAF}-1)} dz\,,
\end{equation}
and 
\begin{equation}
 \tau^{\rm DLA}_{\rm LC}(\lambda_{\rm obs},z_{\rm S})
  \approx \frac{\Gamma(1-\beta_{\rm DLA})}
  {\Gamma(1-\beta_{\rm DLA},N_{\rm l}/N_{\rm c})}
  \int_0^{z_{\rm S}} f_{\rm DLA}(z)
  \left\{1 - (N_{\rm c}\sigma_{\rm L})^{\beta_{\rm DLA}-1}
  \left(\frac{1+z_{\rm L}}{1+z}\right)^{\alpha(\beta_{\rm DLA}-1)}
  \right\} dz\,,
\end{equation}
These integrals can be reduced to the following formulae, 
when $\lambda_{\rm obs}>\lambda_{\rm L}$, if we adopt the
fiducial set of the parameters and the photoionization cross section
index $\alpha=3$. For the LAF component, when $z_{\rm S}<1.2$, 
\begin{equation}
 \tau^{\rm LAF}_{\rm LC}(\lambda_{\rm obs},z_{\rm S}) \approx \cases{
  0.325 \left[
  \left(\frac{\lambda_{\rm obs}}{\lambda_{\rm L}}\right)^{1.2}
  - (1+z_{\rm S})^{-0.9}
  \left(\frac{\lambda_{\rm obs}}{\lambda_{\rm L}}\right)^{2.1}
  \right] & ($\lambda_{\rm obs}<\lambda_{\rm L}(1+z_{\rm S})$) \cr
  0 & ($\lambda_{\rm obs}\geq\lambda_{\rm L}(1+z_{\rm S})$) \cr
  }\,,
\end{equation}
when $1.2\leq z_{\rm S} < 4.7$,
\begin{equation}
 \tau^{\rm LAF}_{\rm LC}(\lambda_{\rm obs},z_{\rm S}) \approx \cases{
  2.55\times10^{-2}(1+z_{\rm S})^{1.6} 
  \left(\frac{\lambda_{\rm obs}}{\lambda_{\rm L}}\right)^{2.1} 
  + 0.325\left(\frac{\lambda_{\rm obs}}{\lambda_{\rm L}}\right)^{1.2}
  -0.250\left(\frac{\lambda_{\rm obs}}{\lambda_{\rm L}}\right)^{2.1} &
  ($\lambda_{\rm obs} < 2.2 \lambda_{\rm L}$) \cr
  2.55\times10^{-2} \left[(1+z_{\rm S})^{1.6}
  \left(\frac{\lambda_{\rm obs}}{\lambda_{\rm L}}\right)^{2.1}
  - \left(\frac{\lambda_{\rm obs}}{\lambda_{\rm L}}\right)^{3.7}
  \right] & ($2.2 \lambda_{\rm L} \leq \lambda_{\rm obs}
  <\lambda_{\rm L}(1+z_{\rm S})$) \cr
  0 & ($\lambda_{\rm obs}\geq\lambda_{\rm L}(1+z_{\rm S})$) \cr
  }\,,
\end{equation}
and when $z_{\rm S}\geq 4.7$, 
\begin{equation}
 \tau^{\rm LAF}_{\rm LC}(\lambda_{\rm obs},z_{\rm S}) \approx \cases{
  5.22\times10^{-4}(1+z_{\rm S})^{3.4} 
  \left(\frac{\lambda_{\rm obs}}{\lambda_{\rm L}}\right)^{2.1} 
  + 0.325\left(\frac{\lambda_{\rm obs}}{\lambda_{\rm L}}\right)^{1.2}
  -3.14\times10^{-2}\left(\frac{\lambda_{\rm obs}}{\lambda_{\rm L}}\right)^{2.1} &
  ($\lambda_{\rm obs} < 2.2 \lambda_{\rm L}$) \cr
  5.22\times10^{-4}(1+z_{\rm S})^{3.4} 
  \left(\frac{\lambda_{\rm obs}}{\lambda_{\rm L}}\right)^{2.1} 
  + 0.218\left(\frac{\lambda_{\rm obs}}{\lambda_{\rm L}}\right)^{2.1}
  -2.55\times10^{-2}\left(\frac{\lambda_{\rm obs}}{\lambda_{\rm L}}\right)^{3.7} &
  ($2.2 \lambda_{\rm L} \leq \lambda_{\rm obs} < 5.7 \lambda_{\rm L}$) \cr
  5.22\times10^{-4} \left[(1+z_{\rm S})^{3.4}
  \left(\frac{\lambda_{\rm obs}}{\lambda_{\rm L}}\right)^{2.1}
  - \left(\frac{\lambda_{\rm obs}}{\lambda_{\rm L}}\right)^{5.5}
  \right] & ($5.7 \lambda_{\rm L} \leq \lambda_{\rm obs}<\lambda_{\rm L}(1+z_{\rm S})$) \cr
  0 & ($\lambda_{\rm obs}\geq\lambda_{\rm L}(1+z_{\rm S})$) \cr
  }\,.
\end{equation}
For the DLA component, when $z_{\rm S}<2.0$, 
\begin{equation}
 \tau^{\rm DLA}_{\rm LC}(\lambda_{\rm obs},z_{\rm S}) \approx \cases{
  0.211(1+z_{\rm S})^{2.0}
  - 7.66\times10^{-2}(1+z_{\rm S})^{2.3}
  \left(\frac{\lambda_{\rm obs}}{\lambda_{\rm L}}\right)^{-0.3} 
  - 0.135\left(\frac{\lambda_{\rm obs}}{\lambda_{\rm L}}\right)^{2.0}
  & ($\lambda_{\rm obs}<\lambda_{\rm L}(1+z_{\rm S})$) \cr
  0 & ($\lambda_{\rm obs}\geq\lambda_{\rm L}(1+z_{\rm S})$) \cr
  }\,,
\end{equation}
and when $z_{\rm S} \geq 2.0$, 
\begin{equation}
 \tau^{\rm DLA}_{\rm LC}(\lambda_{\rm obs},z_{\rm S}) \approx \cases{
  0.634 + 4.70\times10^{-2}(1+z_{\rm S})^{3.0} 
  - 1.78\times10^{-2}(1+z_{\rm S})^{3.3}
  \left(\frac{\lambda_{\rm obs}}{\lambda_{\rm L}}\right)^{-0.3} \cr ~~~~
  - 0.135\left(\frac{\lambda_{\rm obs}}{\lambda_{\rm L}}\right)^{2.0}
  - 0.291\left(\frac{\lambda_{\rm obs}}{\lambda_{\rm L}}\right)^{-0.3}
  & ($\lambda_{\rm obs}<3.0 \lambda_{\rm L}$) \cr
  4.70\times10^{-2}(1+z_{\rm S})^{3.0} 
  - 1.78\times10^{-2}(1+z_{\rm S})^{3.3}
  \left(\frac{\lambda_{\rm obs}}{\lambda_{\rm L}}\right)^{-0.3} \cr ~~~~
  - 2.92\times10^{-2}\left(\frac{\lambda_{\rm obs}}{\lambda_{\rm L}}\right)^{3.0}
  & ($3.0 \lambda_{\rm L} \leq \lambda_{\rm obs}<\lambda_{\rm L}(1+z_{\rm S})$) \cr
  0 & ($\lambda_{\rm obs}\geq\lambda_{\rm L}(1+z_{\rm S})$) \cr
  }\,.
\end{equation}
Note that these formulae are correct when $\lambda_{\rm obs}>\lambda_{\rm L}$.

\subsection{Validity of the analytic formulae}

Let us confirm the validity of the approximate analytic formulae derived
in the two subsections above. We compare the formulae with the numerical
integration of equation~(1). As a result, Figure~7 shows the difference
of the two optical depths divided by the numerical one by a contour in
the plane of the source redshift and the source rest-frame wavelength. 
We find that the differences are less than a few percent in a large area
when the source redshift is larger than 0.5. In the case of 
$z_{\rm S}<0.5$, the observed wavelength for some rest-frame wavelengths 
in the horizontal axis becomes shorter than the Lyman limit, and
then, the formulae for the Lyman continuum absorption in section 4.2 
become incorrect. As a result,
the difference becomes $>10\%$. For $z_{\rm S}>0.5$, the difference
tends to be relatively large for higher order Lyman series lines which 
the DLA component contributes to. Probably the rectangular shape
approximation in the cross section is not very good for it. 
Nevertheless, the difference is still less than several percent and 8\%
at the most, ensuring the validity of the approximate formulae.

\begin{figure*}
 \begin{center}
  \includegraphics[width=10cm]{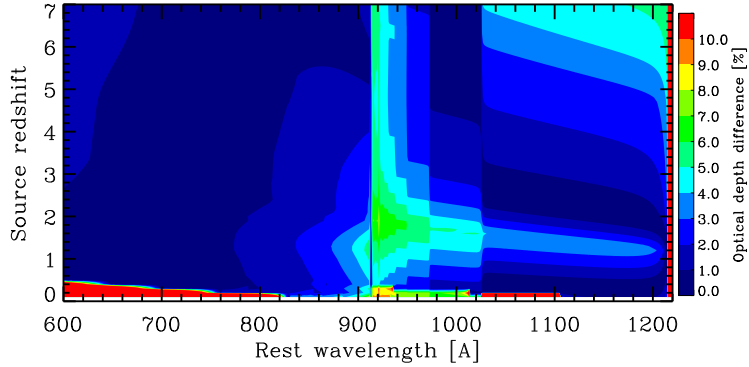}
 \end{center}
 \caption{Fractional difference of the optical depths of the
 numerical integration of equation (1) and the approximate analytic
 formulae presented in sections 4.1 and 4.2. The formulae break down
 at the observed wavelength shorter than the Lyman limit. As a result,
 there appear parts where the difference exceeds 10\% when the source
 redshift $z_{\rm S}<0.52$.}
\end{figure*}

\section{Discussion}

We here compare the attenuation magnitudes through some broad-band
filters for the three models discussed in this paper, quantify the
difference, and discuss the effect on the drop-out technique and
the photometric redshift (hereafter photo-$z$) estimation. 

Suppose the flux density observed through the IGM at the wavelength
$\lambda$ is expressed as 
$F_\lambda^{\rm obs}=F_\lambda^{\rm em}T_\lambda^{\rm IGM}$, where 
$F_\lambda^{\rm em}$ is the emitted flux density at the proximity of a
cosmological object and $T_\lambda^{\rm IGM}$ is the IGM transmission.
We assume a simple power law spectrum for $F_\lambda^{\rm em}$ with a
rest-frame ultra-violet index $\beta_{\rm UV}$: 
$F_\lambda^{\rm em} \propto \lambda^{\beta_{\rm UV}}$. A band magnitude
for using a photon-counting detector like CCDs is defined by
$m=-2.5\log_{10} F + C_0$ with $F=\int (F_\nu t_\nu/h\nu) d\nu / \int
(t_\nu/h\nu) d\nu=\int F_\lambda t_\lambda (\lambda/c) d\lambda / \int
(t_\lambda / \lambda) d\lambda$, where $t_\nu=t_\lambda$ is the total
(including the filter, detector, telescope and instrument optics, 
and atmosphere) efficiency of the band, and $C_0$ is the magnitude 
zero point. Thus, we can express the IGM attenuation through a band
filter as
\begin{equation}
 \Delta m_{\rm IGM} = -2.5 \log_{10} \left(
  \frac{\int F_\lambda^{\rm obs} t_\lambda \lambda d\lambda}
  {\int F_\lambda^{\rm em} t_\lambda \lambda d\lambda} \right)
 = -2.5 \log_{10} \left(
  \frac{\int \lambda^{\beta_{\rm UV}+1} T_\lambda^{\rm IGM} t_\lambda 
  d\lambda}{\int \lambda^{\beta_{\rm UV}+1} t_\lambda d\lambda} \right)\,.
\end{equation}

Figure~8 shows the IGM attenuation magnitudes through 6 broad-band
filters as a function of the source redshift in the case of 
$\beta_{\rm UV}=-2.0$, a flat continuum in $F_\nu$ unit usually
observed in high-$z$ star-forming galaxies \citep[e.g.,][]{sha03}. We 
note here that the variation of $\beta_{\rm UV}$ from $-3$ to 0 
\citep[e.g.,][]{bou09} has a negligible effect on the attenuation
magnitudes. The solid, dotted, and dashed lines are the models of this
work, II08, and M95, respectively. The attenuation magnitudes shown in
the figure are determined mainly by Ly$\alpha$ and Ly$\beta$
absorptions. Then, the difference seems to be small as expected from the
small difference of the mean transmission curves among the three models
shown in Figure~4. In fact, however, the vertical difference at a fixed
source redshift reaches more than 1 mag between this work and the M95
model, while the horizontal difference is as small as about $<0.2$,
except for the deviation of the M95 model at $z_{\rm S}>5.5$ owing to
the lack of the rapid evolution of the Ly$\alpha$ optical depth included
in the other two models. The thin (coloured) solid lines are the results
from the analytic formulae for the new model presented in the previous
section. We find an excellent agreement with the numerical integrations.

\begin{figure}
 \begin{center}
  \includegraphics[width=7cm]{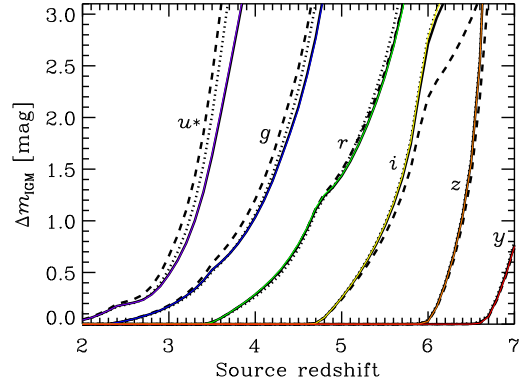}
 \end{center}
 \caption{IGM attenuation magnitude through broad-band filters, the
 Canada-France-Hawaii Telescope/Mega-cam $u^*$, and the Subaru/Hyper
 Suprime-Cam $g$, $r$, $i$, $z$ and $y$, as a function of the source
 redshift. The solid, dotted, and dashed lines are the models of this
 work, Inoue \& Iwata (2008), and Madau (1995), respectively. The thin
 (coloured) solid lines are the cases using the analytic formulae for
 this work. The object spectrum with the ultra-violet spectral index
 $\beta_{\rm UV}=-2.0$ is assumed.}
\end{figure}

\begin{figure}
 \begin{center}
  \includegraphics[width=7cm]{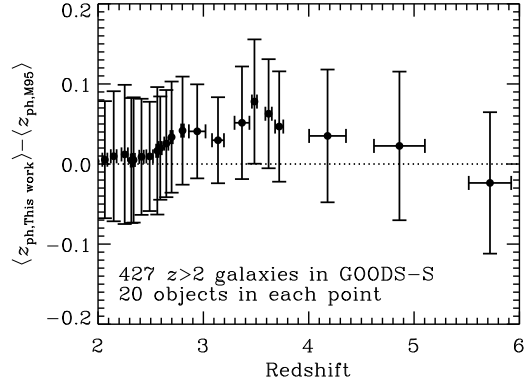}
 \end{center}
 \caption{Difference of the means of the photometric redshift
 estimations assuming the IGM model of this work relative to those
 assuming the Madau (1995) model. The sample is 427 galaxies with
 spectroscopic redshift larger than 2 in the GOODS-S field and is
 divided into bins 20 objects each. Along the horizontal axis, the
 points and error-bars show the mean and the standard deviation of the
 spectroscopic redshifts in each bin. See the text for the vertical
 error-bars.}
\end{figure}

Although the horizontal difference at a certain amount of the
attenuation magnitudes among the three models is small, there is a
difference which would affect the drop-out technique and the 
photo-$z$ estimation. The drop-out threshold is usually 
$\Delta m_{\rm IGM}\simeq1$ mag. The source redshift reaching the
threshold is different from the models. For example, the redshifts in
the M95 model are about 0.2 smaller than those of this work at 
$z_{\rm S}\simeq3$--4 but are about 0.1 larger at $z_{\rm
S}\simeq6$. These difference would result in systematically lower or
higher photo-$z$ solutions with the M95 model than with the new
model of this paper. 
To check this expectation, we ran a photo-$z$ code developed by
\cite{tan13a,tan13b} adopting two IGM models of this paper and M95. 
The sample is the galaxies with spectroscopic redshifts and
photometry of VLT/VIMOS $U$, HST/ACS F435W, F606W, F775W, F814W, F850LP,
HST/WFC3 F105W, F125W, F160W, VLT/ISAAC $K_{\rm s}$, Spitzer/IRAC 
{\it Ch1} and {\it Ch2} in the GOODS-S field \citep{guo13}. We
collected spectroscopic redshifts from the literature
\citep{lef05,mig05,van08,pop09,bal10} and cross-matched with the
photometric objects within 1 arcsec. We use secure redshifts only in the
analysis here. The photo-$z$ code assumes the stellar population
synthesis model by \cite{bru03} with solar and sub-solar metallicity
models ($Z=0.02$, 0.008, and 0.004), the Chabrier initial mass function
between 0.1 and 100 $\rm M_\odot$ \citep{cha03}, exponentially declining
star formation history, the Calzetti attenuation law \citep{cal00}, and
the emission line model by \cite{ino11a} with the Lyman continuum escape
fraction of zero. The Ly$\alpha$ emission line is reduced by a factor of
0.1 to account for the attenuation through the interstellar medium of
galaxies. The metallicity, age, exponential time-scale of the history,
attenuation amount, and redshift are free parameters determined by a
$\chi^2$ minimization technique. We compare the photo-$z$s for the two
IGM models in Figure~9. We divided the sample galaxies into bins 20
objects each and calculated the difference of the means of the
photo-$z$s in each bin. The vertical error-bars are estimated by 
$\sqrt{\sum_i(\sigma_{\langle z_{{\rm ph},i}\rangle}^2/n+\delta_{z_{{\rm ph},i}}^2/n)}$,
where $i$ indicates the two IGM models (this work and M95), 
$\sigma_{\langle z_{{\rm ph},i}\rangle}$ is the standard deviation
of the photo-$z$s in each bin, $\delta_{z_{{\rm ph},i}}$ 
is the mean of photo-$z$ uncertainties of the sample galaxies in each
bin, and $n=20$ is the number of the sample galaxies in each bin. The 
first term is the standard error of the mean and the second term is the
error in the mean propagated from the uncertainty of the individual
photo-$z$. As found in Figure~9, the difference of the means of
photo-$z$s is too small to be detected in the sample adopted, while we 
may find the expected trend of photo-$z$s for the new IGM model larger
(or smaller) than those for the M95 model at $z\simeq3$--5 ($z>5.5$).
The marginal difference of $\approx0.05$ at $z\simeq3.5$ is much smaller
than that expected from Figure~8. This is probably because we used not
only the drop-out band but also all bands available in the photo-$z$
estimation. As a result, the drop-out feature has a lower weight in the
photo-$z$ determination. However, all the available bands should be used
in order to constrain intrinsic shapes of the spectral energy
distribution below Ly$\alpha$ to characterize the IGM effect on
photo-$z$. We would detect the IGM model difference securely if we had
a ten times larger number of the sample galaxies at $z>3$.

\begin{figure}
 \begin{center}
  \includegraphics[width=7cm]{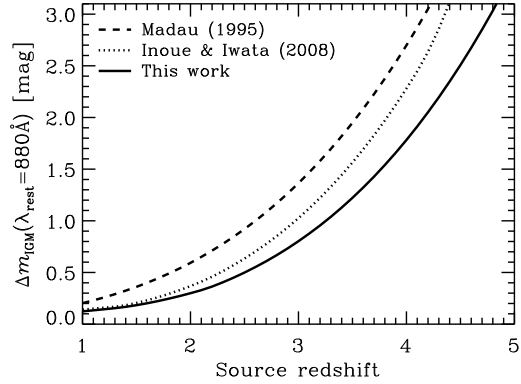}
 \end{center}
 \caption{IGM attenuation magnitude at the rest-frame 880 \AA\ as a
 function of the source redshift. The solid, dotted, and dashed lines
 are the models of this work, Inoue \& Iwata (2008), and Madau (1995),
 respectively.}
\end{figure}

Finally, we examine the mean IGM attenuation magnitude at a Lyman
continuum wavelength 880 \AA\ in the source rest-frame as a function of
the source redshift in Figure~10. This is motivated by studies for
determining an important parameter controlling the cosmic reionization,
the Lyman continuum escape fraction of galaxies
\citep[e.g.,][]{ino05,iwa09}. In these studies, we need to correct the
IGM attenuation against the observed Lyman continuum of galaxies. As
found in Figure~10, the difference among the three models discussed in
this paper is significant; the new model predicts the least attenuation
which is 0.5--1 mag smaller than the M95 model at $z_{\rm S}=3$--4. This
is consistent with those found in Figures~4 and 6. Note that this
less attenuation against the Lyman continuum comes from the recent
updates of the occurrence rate of LLSs discussed in section 2.2 and the
measurements of the mean-free-path discussed in section 3.2. Therefore,
using the M95 model causes a significant overcorrection of the observed
Lyman continuum and results in an overestimation of the escape
fraction. On the other hand, the Lyman continuum absorption is mainly
caused by LLSs which are relatively rare to have on a line-of-sight. As
a result, a large fluctuation of the attenuation amount among many
lines-of-sight is expected. Therefore, a Monte-Carlo simulation is
required to model the stochasticity as done in II08. This point would be
investigated in our next work.

\section*{Acknowledgments}

We would like to thank the referee, J. Xavier Prochaska, for
constructive comments useful to improve this manuscript.
A.K.I. and I.S. are supported by JSPS KAKENHI Grant Number 23684010,
I.I. is supported by JSPS KAKENHI Grant Number 24244018, and M.T. is
supported by JSPS KAKENHI Grant Number 23740144.

\appendix

\section{Analytic integration of the column density distribution}

We have adopted in this paper a function similar to the Schechter
function for the column density distribution of the IGM absorbers as 
\begin{equation}
 g_i(N_{\rm HI}) = B_i {N_{\rm HI}}^{-\beta_i} e^{-N_{\rm HI}/N_{\rm c}}\,,
\end{equation}
where $i$ is either the LAF or DLA components. In this appendix, we
present analytic functions of some integrals of $g_i$.

\subsection{Normalization factor}

The normalization of the column density distribution is set to be 
\begin{equation}
 \int_{N_{\rm l}}^{N_{\rm u}} g_i(N_{\rm HI}) dN_{\rm HI} = 1\,.
\end{equation}
The normalization factor, $B_i$, is then given by
\begin{equation}
 {B_i}^{-1} = \int_{N_{\rm l}}^{N_{\rm u}} 
  {N_{\rm HI}}^{-\beta_i} e^{-N_{\rm HI}/N_{\rm c}} dN_{\rm HI} \,.
\end{equation}
Substituting $x=N_{\rm HI}/N_{\rm c}$ for $N_{\rm HI}$, the integral is
reduced to 
\begin{equation}
 {B_i}^{-1} = {N_{\rm c}}^{1-\beta_i} \int_{x_{\rm l}}^{x_{\rm u}} 
  x^{-\beta_i} e^{-x} dx\,,
\end{equation}
where $x_{\rm l}=N_{\rm l}/N_{\rm c}$ and $x_{\rm u}=N_{\rm u}/N_{\rm c}$.
For the DLA component, we adopt $\beta_{\rm DLA}=0.9$. In this case, we
can obtain the normalization approximately as  
\begin{equation}
 B_{\rm DLA} \approx \frac{{N_{\rm c}}^{\beta_{\rm DLA}-1}} 
  {\Gamma(1-\beta_{\rm DLA},N_{\rm l}/N_{\rm c})}\,,
\end{equation}
where $\Gamma(1-\beta_{\rm DLA},N_{\rm l}/N_{\rm c})$ is an incomplete
Gamma function. We have omit the term 
$\Gamma(1-\beta_{\rm DLA},N_{\rm u}/N_{\rm c})$. On the other hand, we
adopt $\beta_{\rm LAF}=1.7$ for the LAF component. By the method of
integration by parts, equation (A4) can be reduced to 
\begin{equation}
 {B_i}^{-1}=\left[\frac{{N_{\rm HI}}^{1-\beta_i}e^{-N_{\rm HI}/N_{\rm c}}} 
	     {1-\beta_i}\right]_{N_{\rm l}}^{N_{\rm u}} 
 + \frac{{N_{\rm c}}^{1-\beta_i}}{1-\beta_i}
 \int_{x_{\rm l}}^{x_{\rm u}} x^{1-\beta_i}e^{-x} dx\,.
\end{equation}
Since the second term of the right hand side is negligible relative to
the first term for the LAF component, we can obtain the normalization
approximately as
\begin{equation}
 B_{\rm LAF} \approx (\beta_{\rm LAF}-1) {N_{\rm l}}^{\beta_{\rm LAF}-1}\,,
\end{equation}
which is the same as the case of a single power-law distribution function.

\subsection{Integration for the mean optical depth}

In order to perform the integration of equation (1) analytically, we
should consider an approximation of the single absorber optical depth, 
$\tau_{\rm abs}$. If we approximate Lyman series line cross section
profiles to be a narrow rectangular shape, we may treat each line
optical depth and the Lyman continuum optical depth occurring at an 
observed wavelength $\lambda_{\rm obs}$ separately because different
absorbers at different redshifts produce them (see also section 4). 
Then, equation (1) can be reduced to
\begin{equation}
 \langle \tau^{\rm IGM}_{\lambda_{\rm obs}} (z_{\rm S}) \rangle 
  \approx \sum_i \sum_j \int_0^{z_{\rm S}} f_i(z) 
  \int_0^\infty g_i(N_{\rm HI}) (1-e^{-\tau_{{\rm abs},j}}) dN_{\rm HI}
  dz \,,
\end{equation}
where $i$ is either the LAF or DLA components and the optical depth for
$j$th line (including the Lyman continuum absorption) by a single
absorber can be expressed as 
$\tau_{{\rm abs},j}\approx N_{\rm HI} \sigma_j \eta_j(z)$ with
$\sigma_j$ being the $j$th line center cross section (and including 
the Lyman limit cross section $\sigma_{\rm L}$) and 
\begin{equation}
 \eta_j \approx \cases{
  \left(\frac{1+z_{\rm L}}{1+z}\right)^\alpha 
  & (for Lyman continuum absorption) \cr
  1 & (for $j$th Lyman series line absorption) \cr
  }\,,
\end{equation}
where $1+z_{\rm L}=\lambda_{\rm obs}/\lambda_{\rm L}$. The power index
$\alpha\approx3$ \citep{ost89}. If we denote the column density
integration as $I_{i,j}(z)$, it is 
\begin{equation}
 I_{i,j}(z) = \int_0^\infty B_i {N_{\rm HI}}^{-\beta_i} e^{-N_{\rm HI}/N_{\rm c}}
  \{1-e^{-N_{\rm HI} \sigma_j \eta_j(z)}\} dN_{\rm HI}\,.
\end{equation}
Substituting $\tau_j=N_{\rm HI}\sigma_j$ for $N_{\rm HI}$, equation
(A10) can be reduced to
\begin{equation}
 I_{i,j}(z) = B_i {\sigma_j}^{\beta_i-1}
  \int_0^\infty {\tau_j}^{-\beta_i} e^{-\tau_j/\tau_{\rm c}}
  \{1-e^{-\tau_j \eta_j(z)} \} d\tau_j\,
\end{equation}
where $\tau_{\rm c}=N_{\rm c}\sigma_j$. This is analytically integrable
and we obtain for the case of $\beta_i\neq1$ 
\begin{equation}
 I_{i,j}(z) = B_i {\sigma_j}^{\beta_i-1} \Gamma(1-\beta_i) 
  {\tau_{\rm c}}^{1-\beta_i}
  \{1-\left[1+{\tau_{\rm c}}\eta(z)\right]^{\beta_i-1}\}\,,
\end{equation}
where $\Gamma(1-\beta_i)=\Gamma(2-\beta_i)/(1-\beta_i)$ is the Gamma
function. Applying the normalization $B_i$ obtained in appendix A1 and 
$\tau_{\rm c}\gg1$ (and $\eta(z)\sim O(1)$), we finally obtain  
\begin{equation}
 I_{{\rm LAF},j}(z) \approx \Gamma(2-\beta_{\rm LAF}) 
  (N_{\rm l} \sigma_j \eta_j(z))^{\beta_{\rm LAF}-1} \,,
\end{equation}
for the LAF component and 
\begin{equation}
 I_{{\rm DLA},j}(z) \approx \frac{\Gamma(1-\beta_{\rm DLA})}
  {\Gamma(1-\beta_{\rm DLA},N_{\rm l}/N_{\rm c})}
  \{1-(N_{\rm c}\sigma_j\eta_j)^{\beta_{\rm DLA}-1}\} \,,
\end{equation}
for the DLA component. Note that $\beta_{\rm LAF}-1>0$ but 
$\beta_{\rm DLA}-1<0$ for the fiducial set of the parameters in this
paper (see Table~1).

\label{lastpage}


\begin{thebibliography}{99}

\bibitem[Altay et al.(2011)]{alt11}
Altay, G., Theuns, T., Schaye, J., Crighton, N. H. M., Dalla Vecchia,
C., 2011, ApJ, 737, L37

\bibitem[Bahcall \& Peebles(1969)]{bah69}
Bahcall, J. N., Peebles, P. J. E., 1969, ApJ, 156, L7

\bibitem[Balestra et al.(2010)]{bal10} 
Balestra, I., Mainieri, V., Popesso, P., Dickinson, M., Nonino, M., 
Rosati, P., Teimoorinia, H., Vanzella, E., et al., 2010, A\&A, 512, 12

\bibitem[Becker et al.(2013)]{bec13}
Becker, G. D., Hewett, P. C., Worseck, G., Prochaska, J. X., 2013,
MNRAS, 436, 1023

\bibitem[Bershady et al.(1999)]{ber99}
Bershady M. A., Charlton J. C., Geoffroy J. M., 1999, ApJ, 518, 103

\bibitem[Bouwens et al.(2009)]{bou09}
Bouwens, R. J., Illingworth, G. D., Franx, M., Chary, R.-R.,
Meurer, G. R., Conselice, C. J., Ford, H., Giavalisco, M., 
van Dokkum, P., 2009, ApJ, 705, 936

\bibitem[Bruzual \& Charlot(2003)]{bru03}
Bruzual, G., Charlot, S., 2003, MNRAS, 344, 1000

\bibitem[Calzetti et al.(2000)]{cal00}
Calzetti, D., Armus, L., Bohlin, R. C., Kinney, A. L., 
Koornneef, J., Storchi-Bergmann, T., 2000, ApJ, 533, 682

\bibitem[Cen et al.(1994)]{cen94}
Cen, R., Miralda-Escud{\'e}, J., Ostriker, J. P., Rauch, M., 
1994, ApJ, 437, L9

\bibitem[Chabrier(2003)]{cha03}
Chabrier, G., 2003, PASP, 115, 763

\bibitem[Corbelli et al.(2001)]{cor01}
Corbelli, E., Salpeter, E. E., Bandiera, R., 2001, ApJ, 550, 26

\bibitem[Corbelli \& Bandiera(2002)]{cor02}
Corbelli, E., Bandiera, R., 2002, ApJ, 567, 712

\bibitem[Cox(2000)]{cox00}
Cox, A. N., 2000, Allen's Astrophysical Quantities, 4th ed., AIP press,
Springer, New York

\bibitem[Dav{\'e} et al.(1999)]{dav99}
Dav{\'e} R., Hernquist L., Katz N., Weinberg D. H., 1999, ApJ, 511, 521

\bibitem[Fan et al.(2006)]{fan06}
Fan, X., Strauss, M. A., Becker, R. H., White, R. L., Gunn, J. E., 
Knapp, G. R., Richards, G. T., Schneider, D. P., Brinkmann, J., 
Fukugita, M., 2006, AJ, 132, 117

\bibitem[Faucher-Gigu{\`e}re et al.(2008)]{fau08}
Faucher-Gigu{\`e}re, C.-A., Prochaska, J. X., Lidz, A., Hernquist, L.,
Zaldarriaga, M., 2008, ApJ, 681, 831

\bibitem[Fumagalli et al.(2013)]{fum13}
Fumagalli, M., O'Meara, J. M., Prochaska, J. X., Worseck, G., 
2013, ApJ, 775, 78

\bibitem[Gunn \& Peterson(1965)]{gun65}
Gunn, J. E., Peterson, B. A., 1965, ApJ, 142, 1633

\bibitem[Guo et al.(2013)]{guo13}
Guo, Y., Ferguson, H. C., Giavalisco, M., Barro, G., Willner, S. P., 
Ashby, M. L. N., Dahlen, T., Donley, J. L., et al., 2013, ApJS, 207, 24

\bibitem[Haehnelt et al.(1998)]{hae98}
Haehnelt, M. G., Steinmetx, M., Rauch, M., 
1998, ApJ, 495, 647

\bibitem[Harrison et al.(2011)]{har11}
Harrison, C. M., Meiksin, A., Stock, D., 2011, arXiv:1105.6208

\bibitem[Hui \& Rutledge(1999)]{hui99}
Hui L., Rutledge R. E., 1999, ApJ, 517, 541

\bibitem[Inoue et al.(2005)]{ino05}
Inoue, A. K., Iwata, I., Deharveng, J.-M., Buat, V., Burgarella, D.,
2005, A\&A, 435, 471

\bibitem[Inoue \& Iwata(2008)]{ino08}
Inoue, A. K., Iwata, I., 2008, MNRAS, 387, 1681 (II08)

\bibitem[Inoue(2011)]{ino11a}
Inoue, A. K., 2011, MNRAS, 415, 2920

\bibitem[Inoue et al.(2011)]{ino11}
Inoue, A. K., Kousai, K., Iwata, I., Matsuda, Y., Nakamura, E., Horie,
M., Hayashino, T., Tapken, C., et al., 2011, MNRAS, 411, 2336

\bibitem[Iwata et al.(2009)]{iwa09}
Iwata, I., Inoue, A. K., Matsuda, Y., Furusawa, H., Hayashino, T.,
Kousai, K., Akiyama, M., Yamada, T., et al.,
2009, ApJ, 692, 1287

\bibitem[Janknecht et al.(2006)]{jan06}
Janknecht, E., Reimers, D., Lopez, S., Tytler, D., 2006, A\&A, 458, 427

\bibitem[Kim et al.(2001)]{kim01}
Kim, T.-S., Cristiani S., D'Odorico S., 2001, A\&A, 373, 757

\bibitem[Kim et al.(2013)]{kim13}
Kim, T.-S., Partl, A. M., Carswell, R. F., M{\"u}ller, V., 
2013, A\&A, 552, 77 

\bibitem[Kirkman et al.(2007)]{kir07}
Kirkman, D., Tytler, D., Lubin, D., Charlton, J., 2007, MNRAS, 376, 1227

\bibitem[Le F{\`e}vre et al.(2005)]{lef05} 
Le F{\`e}vre, O., Vettolani, G., Garilli, B., Tresse, L., Bottini, D., 
Le Brun, V., Maccagni, D., Picat, J. P., et al., 2005, A\&A, 439, 845

\bibitem[Madau(1995)]{mad95}
Madau, P., 1995, ApJ, 441, 18 (M95)

\bibitem[Madau et al.(1996)]{mad96}
Madau, P., Ferguson, H. C., Dickinson, M. E., Giavalisco, M., Steidel, 
C. C., Fruchter, A., 1996, MNRAS, 283, 1388

\bibitem[Meiksin(2006)]{mei06}
Meiksin, A., 2006, MNRAS, 365, 807

\bibitem[Mignoli et al.(2005)]{mig05} 
Mignoli M., et al., 2005, A\&A, 437, 883

\bibitem[M{\o}ller \& Jakobsen(1990)]{mol90}
M{\o}ller, P., Jakobsen, P., 1990, A\&A, 228, 299

\bibitem[O'Meara et al.(2013)]{ome13}
O'Meara, J. M., Prochaska, J. X., Worseck, G., Chen, H.-W., Madau, P., 
2013, ApJ, 765, 137

\bibitem[Osterbrock(1989)]{ost89}
Osterbrock D. P., 1989, in Astrophysics of Gaseous Nebulae and Active
Galactic Nuclei. University Science Books, Mill Valley, CA

\bibitem[Overzier et al.(2013)]{ove13}
Overzier, R., Lemson, G., Angulo, R. E., Bertin, E., Blaizot, J., 
Henriques, B. M. B., Marleau, G.-D., White, S. D. M., 2013, MNRAS, 428, 778

\bibitem[Paresce et al.(1980)]{par80}
Paresce F., McKee C. F., Bowyer S., 1980, ApJ, 240, 387

\bibitem[P{\'e}roux et al.(2005)]{per05}
P{\'e}roux, C., Dessauges-Zavadsky, M., D'Odorico, S., Kim, T.-S.,
McMahon, R. G., 2005, MNRAS, 363, 479

\bibitem[Popesso et al.(2009)]{pop09} 
Popesso P., Dickinson, M., Nonino, M., Vanzella, E., Daddi, E., 
Fosbury, R. A. E., Kuntschner, H., Mainieri, V., et al., 2009, A\&A, 494, 443

\bibitem[Prochaska et al.(2005)]{pro05}
Prochaska, J. X., Herbert-Fort, S., Wolfe, A. M., 2005, ApJ, 635, 123

\bibitem[Prochaska \& Wolfe(2009)]{pro09a}
Prochaska, J. X., Wolfe, A. M., 2009, ApJ, 696, 1543

\bibitem[Prochaska et al.(2009)]{pro09b}
Prochaska, J. X., Worseck, G., O'Meara, J. M., 2009, ApJ, 705, L113 (PWO)

\bibitem[Prochaska et al.(2010)]{pro10}
Prochaska, J. X., O'Meara, J. M., Worseck, G., 2010, ApJ, 718, 392

\bibitem[Prochaska et al.(2014)]{pro14}
Prochaska, J. X., Madau, P., O'Meara, J. M., Fumagalli, M., 2014, MNRAS,
438, 476

\bibitem[Rahmati et al.(2013)]{rah13}
Rahmati, A., Pawlik, A. H., Rai{\v c}evi{\`c}, Schaye, J., 
2013, MNRAS, 430, 2427

\bibitem[Rao et al.(2006)]{rao06}
Rao, S. M., Turnshek, D. A., Nestor, D. B., 2006, ApJ, 636, 610

\bibitem[Rauch(1998)]{rau98}
Rauch, M., 1998, ARA\&A, 36, 267

\bibitem[Ribaudo et al.(2011)]{rib11}
Ribaudo, J., Lehner, N., Howk, J. C., 2011, ApJ, 736, 42

\bibitem[Rudie et al.(2013)]{rud13}
Rudie, G. C., Steidel, C. C., Shapley, A. E., Pettini, M., 
2013, ApJ, 769, 146

\bibitem[Shapley et al.(2003)]{sha03}
Shapley, A. E., Steidel, C. C., Pettini, M., Adelberger, K. L.,
2003, ApJ, 588, 65

\bibitem[Slosar et al.(2011)]{slo11}
Slosar, A., Font-Ribera, A., Pieri, M. M., Rich, J., Le Goff, J.-M.,
Aubourg, R., Brinkmann, J., Busca, N., et al., 2011, Journal of
Cosmology and Astroparticle Physics, 9, 1

\bibitem[Songaila \& Cowie(2010)]{son10}
Songaila, A., Cowie, L. L., 2010, ApJ, 721, 1448

\bibitem[Steidel et al.(1995)]{ste95}
Steidel, C. C., Pettini, M., Hamilton, D., 1995, AJ, 110, 2519

\bibitem[Tanaka et al.(2013a)]{tan13a}
Tanaka, M., Finoguenov, A., Mirkazemi, M., Wilman, D. J., Mulchaey,
J. S., Ueda, Y., Xue, Y., Brandt, W. N., Cappelluti, N.,
2013a, PASJ, 65, 17

\bibitem[Tanaka et al.(2013b)]{tan13b}
Tanaka, M., Toft, S., Marchesini, D., Zirm, A., De Breuck, C., 
Kodama, T., Koyama, Y., Kurk, J., Tanaka, I., 
2013b, ApJ, 772, 113

\bibitem[Telfer et al.(2002)]{tel02}
Telfer, R. C., Zheng, W., Kriss, G. A., Davidsen, A. F., 
2002, ApJ, 565, 773

\bibitem[Tepper-Garc{\'i}a(2006)]{tep06}
Tepper-Garc{\'i}a T., 2006, MNRAS, 369, 2025

\bibitem[Tepper-Garc{\'i}a \& Fritze(2008)]{tep08}
Tepper-Garc{\'i}a T., Fritze U., 2008, MNRAS, 383, 1671

\bibitem[Tytler(1987)]{tyt87}
Tytler, D., 1987, ApJ, 321, 49

\bibitem[Vanzella et al.(2008)]{van08}
Vanzella E., Cristiani, S., Dickinson, M., Giavalisco, M., 
Kuntschner, H., Haase, J., Nonino, M., Rosati, P., et al., 
2008, A\&A, 478, 83

\bibitem[Wiese et al.(1966)]{wie66}
Wiese, W. L., Smith, M. W., Glennon, B. M., 1966, Atomic Transition
Probabilities, 1, US Department of Commerce, National Bureau of
Standards, Washington

\bibitem[Weymann et al.(1998)]{wey98}
Weymann, R. J., Jannuzi, B. T., Lu, L., Bahcall, J. N., Bergeron, J., 
Boksenberg, A., Hartig, G. F., Kirhakos, S., et al., 
1998, ApJ, 506, 1

\bibitem[Worseck et al.(2014)]{wor14}
Worseck, G., Prochaska, J. X., O'Meara, J. M., Becker, G. D., Ellison,
S. L., Lopez, S., Meiksin, A., M{\'e}nard, B., et al.,
2014, MNRAS, submitted (arXiv:1402.4154)

\bibitem[Yoshii \& Peterson(1994)]{yos94}
Yoshii, Y., Peterson, B. A., 1994, ApJ, 436, 551

\bibitem[Zuo(1993)]{zuo93}
Zuo, L., 1993, A\&A, 278, 343

\end{thebibliography}
\end{document}